\documentclass[runningheads]{svmult}

\usepackage{makeidx}   
\usepackage{graphicx}  
\usepackage{subeqnar}  
\usepackage{multicol}  
\usepackage{physprbb}  
\makeindex             

\usepackage{epsfig}
\begin{document}
\title*{Superfluidity in relativistic neutron stars}
\toctitle{Superfluidity in relativistic neutron stars}
%
%
\titlerunning{Superfluidity in relativistic neutron stars}
%
\author{David Langlois\inst{1,2}}
\authorrunning{David Langlois}
%
%
\institute{{\it D\'epartement d'Astrophysique Relativiste et de Cosmologie}, 
Observatoire de Paris-Meudon, 92195 Meudon, France
\and {\it Institut d'Astrophysique de Paris} , 
98bis Boulevard Arago, 75014 Paris, France}

\maketitle              

\begin{abstract}
The purpose of these notes is to give
a brief review  of superfluidity in neutron stars. After a short 
presentation explaining why and how superfluidity is expected in the crust 
and core of neutron stars, consequences on thermal evolution and rotational 
dynamics are discussed. The second part summarizes  a formalism that has been 
recently developped to describe the hydrodynamics of superfluids or 
superconductors in the framework of general relativity. As an application, 
one can compute the oscillations of a two-component relativistic neutron star.

\end{abstract}

\def\beq{\begin{equation}}
\def\eeq{\end{equation}}

\def\Om{{\Omega}}

\def\epp{\epsilon^{\prime}}
\def\vep{\varepsilon}
\def\ra{\rightarrow}
\def\ppg{\pi^+\pi^-\gamma}
\def\vp{{\bf p}}
\def\ko{K^0}
\def\kb{\bar{K^0}}
\def\al{\alpha}
\def\ab{\bar{\alpha}}
\def\be{\begin{equation}}
\def\ee{\end{equation}}
\def\bea{\begin{eqnarray}}
\def\eea{\end{eqnarray}}
\def\CPbar{\hbox{{\rm CP}\hskip-1.80em{/}}}
\def\d{\delta}
\def\fNdrag{{f_n}_{\!_{\perp}}}
\def\drag{{\cal C_{\rm r}}}
\def\JC{{J_{\rm c}}}
 \def\Si{{\mit \Sigma}}

\section{Introduction}

The present contribution will discuss superfluidity and superconductivity
in a rather extreme environment: neutron stars. 
Neutron stars, dense stars composed mainly of neutrons, were envisaged by 
Landau as 
soon as the neutron was discovered in 1932. In 1934, Baade and Zwicky 
suggested that  supernovae were manifestations of  a transition  from an 
ordinary star to a
very dense  neutron star. In 1967, Hewish and Bell  discovered  the first  
``pulsar'' (pulsating source of radio), which was soon identified as 
a rotating neutron star. The following year, two new pulsars, much 
studied until now, were discovered: the Vela pulsar with a period $P=89 $ ms
and the Crab pulsar with $P=33$ ms. 
Since then, more than one thousand pulsars have been detected. 

Neutron stars
are rather impressive objects. They contain a mass of the order of the solar
mass confined in a radius  $R\sim 10$ km, which implies an average mass density
of the order $\rho\sim 10^{14} {\rm g/cm^3}$ . They can rotate up to 
several hundred times per second. 
Due to these extreme conditions, neutron stars are of interest for various 
branches of physics. First, neutron stars are so dense and so compact that 
their gravitational field is very strong,
\beq
{GM\over c^2 R}\sim 0.2,
\eeq
and can be described correctly 
only with  general gelativity.
Moreover, binary pulsars have been extraordinary ``laboratories'' to test the 
strong field predictions of general relativity, in particular to verify 
(indirectly) the existence of gravitational radiation. 

Second, the magnetic field, typically of the order of $10^{12}$ G, plays 
a very important r\^ole in the neutron star physics, in particular 
in the pulsar emission mechanism (see e.g. \cite{michel}).
Third, neutron stars are unique places where one   can 
find matter in an ultra-dense state. The density is indeed higher than 
the atomic nuclear density. The corresponding pressure therefore 
depends essentially on the strong interactions, and neutron star observations
are potentially a very rich source of information about the behaviour 
of strong interactions at high densities, which, until now, 
 remains very poorly known 
from both theoretical and experimental points of view. 

Finally, neutron star physics has also some connection with the low 
temperature physics. 
Studying the nucleon interactions,  Migdal suggested the 
possibility that neutron star matter become superfluid at sufficiently low
temperature. If the   temperature of neutron stars, typically 
$T\simeq 10^6 K$,  may seem huge in terrestrial standards, 
 it is in fact smaller than the superfluid critical temperature 
evaluated as $T_c \sim 10^9-10^{10}$ K. 

In these notes, we will focus on two aspects of the physics of neutron stars: 
the property of superfluidity with its consequences on the evolution 
of a neutron star; the necessity to consider a neutron star as a 
general relativistic object. Many other interesting aspects of neutron stars 
will thus be left aside and the curious reader will find some useful 
information in e.g. \cite{ST}. 

The plan will be the following: we will begin with some details about the 
superfluid properties of the interior of neutron stars. Observable 
consequences of superfluidity will then be discussed: first, the 
impact on the thermal evolution of neutron stars, then on their  rotational 
dynamics. In a second part, we will summarize 
some elements of a formalism that gives a hydrodynamical description of 
the interior of neutron stars in the framework of general relativity. Finally, 
as an illustration, oscillations of two-component superfluid neutron stars are
condidered.

\section{Superfluidity and superconductivity}

\subsection{Composition of the interior of a neutron star}

Let us start by describing the matter composition inside a neutron star. 
Because there is a strong density gradient from the exterior to the interior 
of the star, the composition changes dramatically with the  radial distance 
from  the center. Let us progress from the surface of the star 
towards its interior. 

Ignoring here the surface ocean, the first layer of a neutron star is the 
{\it outer crust}: it is made of a lattice of nuclei with a gas of 
relativistic degenerate electrons. 
The nuclei are richer and richer in neutrons as the density increases 
(see e.g.  \cite{pr95} for a recent review).
 At some point ($\rho \simeq 4.3\times 10^{11} {\rm g/cm^3}$),
 neutrons begin to leak out of the nuclei. This is called the 
{\it neutron drip} transition 
and this marks the boundary between the outer crust and 
the {\it inner crust}. For the latter, one still has  a lattice of nuclei 
immersed in an electron gas, but in addition there is a liquid  of free 
neutrons. 
The density still increasing, the difference between the neutrons in nuclei 
and the neutrons outside becomes fainter and fainter until the nuclei 
simply dissolve
(at a density $\rho\sim 2\times 10^{14} {\rm g/cm^3}$).
One has reached the limit of the crust and one now enters the realm of the 
core. 

In the {\it outer core}, one will find the coexistence of a gas free neutrons, 
which is by far the main species, with a liquid of  free protons, plus 
a gas of 
electrons (and muons) which ensures charge neutrality. 
Finally the deeper layers of a neutron star, the {\it inner core}, remain 
mysterious. 
Several possibilities can be envisaged, including appearance of hyperons, 
condensation of pions or kaons, or the transition from hadronic to quark 
matter.
\begin{figure}[b]
\begin{center}
\input{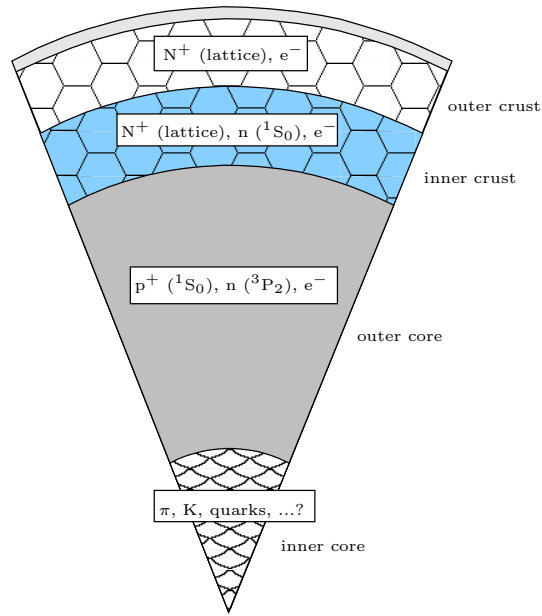}
\end{center}
\caption[]{Schematic view of the internal structure of a neutron star}
\label{eps1}
\end{figure}

\subsection{Energy gaps and critical temperature}

Like electrons in a superconductor, the superfluidity of neutrons in neutron 
stars is due to the pairing of two neutrons near the Fermi surface in 
momentum space, according to the Cooper mechanism (see e.g. \cite{sauls89}). 
In the inner crust, the neutron Cooper pairs are preferably in a state 
${}^1 S_{0}$. 
In the outer core, where the density is much higher, the neutron 
Cooper pairs are of the ${}^3P_{2}$ type 
(see \cite{hgrr70} and \cite{Takatsuka72}). The evaluation of the energy gaps 
give typically values of the order of $1$ MeV. 
 
In the same layer, the protons 
are free and can undergo the same process as neutrons, 
i.e. form Cooper pairs. Being fewer than the neutrons, they are expected 
to condense into 
a    ${}^1 S_{0}$ state \cite{ccy72}. 
Note that hyperons that may exist in the inner core are also expected to be
superfluid. 

\subsection{Various equations of state}
In order to determine the global structure of neutron stars, one needs the 
effective bulk equation of state of neutron star matter. Inserting this 
equation of state in  Einstein's equations, it is immediate, in the case 
of a {\it non-rotating star} (see subsection 6.1),  to 
 obtain the radial profile of the star, i.e. the radial profile of 
the energy density, of the pressure and of the metric coefficients. 
One can also take into account the rotation of the star.

The main problem however is that the equation of state is unknown at very 
high densities. The reason is the lack of experimental data as well as the 
theoretical and computational challenge to evaluate the interactions of 
high density matter. Therefore there exist many different equations of state 
in the literature (see e.g. \cite{hp00} for a recent review) and the hope 
is to infer from observations of neutron stars some constraints on the 
high density equations of state  (see 
\cite{glendenning} and \cite{lp00}).

\section{Cooling processes in neutron stars}

Born with a temperature of the order $10^{11}-10^{12}$ K, neutron stars 
cool very rapidly to temperatures of less than $10^{10}$ K within minutes.
The subsequent thermal evolution is more uncertain and   could 
 be strongly affected by the presence of superfluidity. 
The mechanism by which  neutron stars cool down is 
 essentially neutrino emission. 
Neutrinos can be produced in two types of processes:
\begin{itemize}
\item Direct URCA processes

They correspond to the simplest beta reactions
\beq
n\rightarrow p+e^-+\bar\nu_e, \quad p+e^-\rightarrow n+\nu_e.
\eeq

\item Modified URCA processes

They correspond to beta reactions with the presence of a second nucleon.
\end{itemize}

Direct URCA ractions are (not surprisingly) more efficient than the modified 
ones but they can occur only if conservation of  energy and momentum is 
satisfied, which implies some constraint  on the species fractions, namely 
that the proton fraction is sufficiently high. If this is not the case, then 
direct URCA processes are completely suppressed and cooling occurs only 
through modified URCA processes and is much slower. 

However, taking into account the existence of superfluidity leads to the 
conclusion that  the direct URCA  processes could be {\it partially}
 suppressed: the neutrino emissivity would be reduced by a factor 
$\exp(-\Delta/kT)$, where $\Delta$ is the superfluid gap energy. The reason 
is that, before the beta reaction takes place, one needs to break the Cooper 
pair. 

The existence of superfluidity therefore enables us to envisage, between 
the two extreme scenarios suggested above of {\it fast cooling} or {\it slow 
cooling}, an intermediate scenario of {\it moderate cooling}, which may 
be testable in the future by the temperature measurements of neutron stars
 (see \cite{ppls00} and 
references therein).

\section{Rotational dynamics of neutron stars: glitches}

Another important consequence of superfluidity concerns the rotational 
dynamics of neutron stars, which can be followed extremely precisely via  
the observation of the radio signal from pulsars.
 The main feature is an extremely stable periodic signal, the pulsar
being analogous to a lighthouse beacon: the radio emission is indeed 
collimated along the magnetic axis which does not coincide with the 
rotation axis.

For isolated pulsars, one observes  a steady, although tiny, 
 increase of the pulse period, i.e. a steady 
decrease of the  angular velocity. This is interpreted as a loss of 
angular momentum  due to the emission of electromagnetic radiation.
Assuming a magnetic dipole, the rotational energy loss is given by 
\beq
\dot E=I\Omega \dot \Omega=-{B^2 R^6\Omega^4\sin^2\theta\over 
6 c^3},
\eeq
where $I$ is the moment of inertia and $\theta$ the angle between the 
magnetic and rotation axes. 

In addition to this slow increase of the 
period, a few pulsars exhibit some rare events named {\it glitches }
at  which the period suddenly decreases, i.e. the 
angular velocity suddenly increases, followed by a slow exponential 
relaxation with a 
typical time scale of the order of days to months.   The  two most famous 
`glitching' pulsars are the Vela and Crab pulsars with period changes 
$\Delta \Omega/\Omega$ of the order of $10^{-6}$ and $10^{-8}$ respectively.

The current explanation for these glitches is the following. It is based 
on the idea that the neutron star interior contains a weakly coupled 
component that is not directly slowed down by the electromagnetic torque
and therefore rotates  slightly faster than  the star crust. 
What is directly observable
is the angular velocity of the crust, and if, for some reason, there is a 
rapid transfer of angular momentum from  the faster component to the 
crust, then one should see a sudden increase  of the pulsar period. 
Of course, a natural candidate for the component rotating faster is the 
{\it neutron superfluid}.

\subsection{The two-fluid model}

The  relaxation after the glitch can also be explained, at 
least qualitatively,  by the 
simple two-component model of Baym et al \cite{BPPR69}. 
One considers two components 
with respective angular velocities $\Omega_c$ and  $\Omega_n$, and respective
moments of inertia, $I_c$ and $I_n$. The first component corresponds to the 
crust and to whatever is strongly coupled to it, whereas the second component
corresponds to the neutron superfluid weakly coupled to the crust.
Note that the pulsar angular velocity $\Omega$ which is the directly observable
quantity can be identified with the crust angular velocity, so that 
$\Omega_c=\Omega$. The evolution 
of the two components is governed by the system 
\begin{eqnarray}
I_c\dot \Omega&=&-{I_c\over \tau_c}(\Omega-\Omega_n)-\alpha,\\
I_n\dot \Omega_n&=&{I_c\over \tau_c}(\Omega-\Omega_n),
\end{eqnarray}
where the first term on the right hand side represents a coupling between 
the two components, coupling characterized by the timescale $\tau_c$.
$\alpha$ corresponds to the electromagnetic torque. One can integrate 
the above system just after a glitch characterized by a sudden jump of the 
crust angular velocity $\Delta\Omega_0\equiv \Omega(t=0^+)-\Omega(t=0^-)$. 
The subsequent evolution is then 
\begin{equation}
\Omega(t)=\Omega_0(t)+\Delta\Omega_0\left(Qe^{-t/\tau}+1-Q\right),
\end{equation}
where $Q$ is the healing parameter and $\Omega_0(t)\equiv 
\Omega_0-(\alpha/I)t$ ($\Omega_0$ being a constant) corresponds to the 
evolution without glitch, and the characteristic relaxation timescale is 
given by 
\beq
\tau ={I_n\over I}\tau_c, 
\eeq
with $I=I_n+I_c$. 

\begin{figure}[b]
\begin{center}
\input{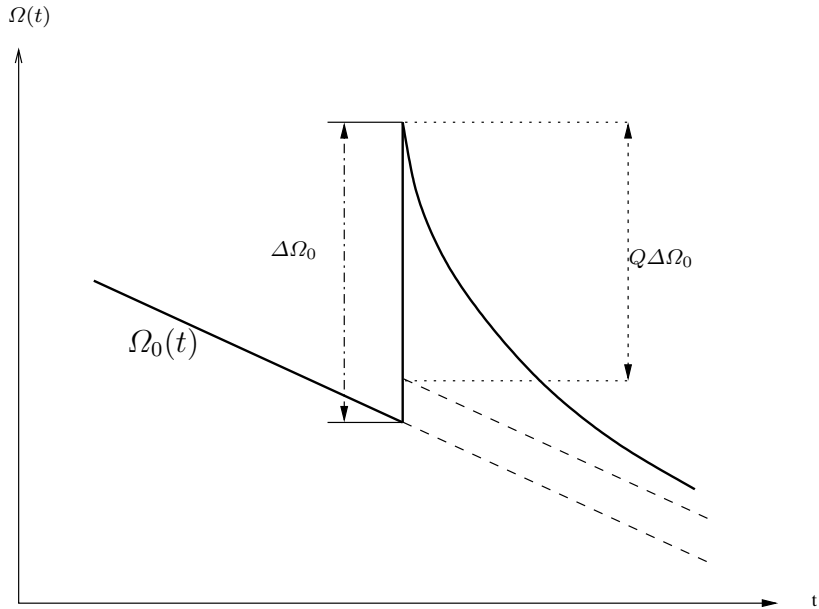}
\end{center}
\caption[]{Evolution of the angular velocity during a glitch}
\label{eps2}
\end{figure}

\subsection{R\^ole of the vortices}
The two-fluid model gives an effective view of the rotational evolution, 
but it is interesting to explore in more details the underlying mechanism 
responsible for the coupling between the two components. This is where the 
vortices enter into play. 

The neutron star at its birth is a rapidly rotating object. With its cooling, 
the temperature reaches the critical temperature under which superfluidity 
appears. However, one essential property of a superfluid  is that 
its flow is {\it irrotational}. 
The way the superfluid solves the contradiction 
between irrotationality and rotation of the star is, like 
superfluid Helium in laboratories,  via  the creation 
of quantized vortices that carry all the angular momentum of the superfluid. 
The superfluid `velocity' (in fact, momentum) being the gradient of a phase,
\beq
\vec p_n=2m_n \vec v_n=\hbar {\vec\nabla}\varphi,
\eeq
 the circulation around any 
closed path is given by
\beq
\kappa\equiv \int_{C}\vec{dl } \vec v_n={\hbar\over 2m_n} (2\pi N),
\eeq
where $N$ is an integer. Superfluid vortices usually correspond to one unit 
($N=1$) of 
quantized circulation $\kappa=h/(2m_n)$. One can then easily compute the 
average density of vortices for a superfluid rotating at uniform angular 
velocity $\Omega_n$:
\beq
n_V=4 {\Om_n m_n\over h}.
\eeq
Putting numbers, one finds
\beq
n_V= 6.3\times 10^3\left ({P\over 1\,{\rm s}}\right)^{-1} 
{\rm vortices\,\,  per\,\,  cm^2} 
\eeq

In the neutron star core, the existence of a proton  superconductor, 
which is believed to be a type II superconductor,  suggests the presence
 of magnetic vortices, or fluxoids, which carry the  magnetic flux 
going through the neutron star. Note that one must be careful to compute 
the total energy of a fluxoid in a {\it rotating} superconducting background
\cite{cpl99}.
Moreover, because of the `entrainment effect' by which the neutron superfluid
momentum is associated with both neutron and proton currents,  
neutron superfluid vortices, in the neutron star core, will carry a magnetic 
flux as well, thus implying a strong coupling between the core superfluid 
and the crust \cite{vs81,als84}.

Now, the vortices play an essential r\^ole for the coupling of the superfluid
to the normal component of the star, because the pure superfluid has no 
viscosity at all and, as such,  can circulate without affecting any normal 
component. But the vortices interact with the normal component. Without 
entering the   details, one can distinguish two types of interaction 
between the vortices and the normal part of a neutron star:
\begin{itemize}
\item {\it Pinning}

This is the situation where a vortex, or a segment of it, is anchored to the 
(inner) crust. The pinning is due to the interaction between the vortex 
and the nucleus, which can be attractive or repulsive depending on the 
matter density \cite{eb88,pvb97}

\item {\it Friction}

This is the case where the vortex, moving through the normal component, 
interacts with it in some dissipative processes, which results into an 
effective friction force (per unit length)  exerted on the vortex by 
the normal component of 
the form 
\beq
\vec F_{\rm drag}={\cal C}(\vec v_V-\vec v_c). \label{drag}
\eeq

\end{itemize}

Let us assume here that we are in the friction case. 
The vortices, interacting with the normal component, are also sensitive 
to the motion of the superfluid through a Magnus force term,
\beq
\vec F_{M}=n_n \vec \kappa\times (\vec v_V-\vec v_n), \label{magnus}
\eeq
which is a force per unit vortex length.
The effective motion of the vortices is obtained by requiring that the 
two forces (\ref{drag}) and (\ref{magnus}) just cancel. This implies that the 
vortices must have a radial motion in addition to the angular motion. One finds
easily that the angular motion is given by
\beq
\Omega_V={c_r^{-1}\Omega_n +c_r \Omega_c\over c_r^{-1}+c_r },
\eeq
which means that the angular velocity of the vortices is simply a weighted 
average of the superfluid and normal angular velocities, the weight 
depending on the dimensionless friction coefficient
\beq
c_r={{\cal C}\over \kappa n_n}.
\eeq
The two extreme cases are: the case where the friction is zero, the vortices 
being then in corotation with the superfluid; the case where the friction 
is huge, the vortices being then in corotation with  the normal component.
As for the radial velocity of the vortices, it is given by 
\beq
v_V^r={\Omega_n-\Omega_c\over c_r^{-1}+c_r }r.
\eeq

One can then compute the effective coupling timescale between the two 
components of the stars, by noting that the evolution of the angular velocity
of the superfluid component is directly given by the radial velocity of 
the vortex array, according to the expression 
\beq
\dot \Omega_n=-2{v_V^r\over r}\Omega_n,
\eeq
simply because the angular momentum of the superfluid is proportional to the 
vortex density. Let us consider now the evolution of the angular velocity 
lag between the two components,
\beq
\omega\equiv \Omega_n-\Omega_c,
\eeq
and let us define the coupling timescale as $\tau_d\equiv |\omega/\dot\omega|$.
Combining the expressions obtained above, one 
finds 
\beq
\tau_d\simeq {c_r+c_r^{-1}\over 2}\left({I_c\over I_c+I_n}\right)\Omega^{-1}.
\eeq

The above Newtonian analysis has been generalized to a general relativistic 
context in \cite{lsc98}. The coupling between the crust and the superfluid 
can also be evaluated when vortices are pinned, in which case vortices can 
move outwards by vortex creep, the temperature being a crucial parameter 
\cite{aaps84a}.

\subsection{Origin of the glitches}
There is no certainty at present on the physical origin of observed glitches. 
The scenario which seems to have attracted the greatest attention is the one 
proposed by Anderson and Itoh \cite{AI75}. In this model, 
 vortices are supposed to be pinned in the crust due to the nucleus-vortex 
interaction. Because the normal component is slowing down, a Magnus force, due
to the superfluid motion relative to the vortices, progressively builds 
up until this force reaches a critical amplitude for which the pinning 
sites simply break. The vortices then can suddenly move outwards, thus 
transfering angular momentum from the superfluid to the crust, which 
explains the sudden spin-up of the crust. 

Other models have been proposed to explain glitches.  For example, a glitch 
could be induced by a sudden heat wave propagating in the star, which would 
increase  the effective coupling between the superfluid and the crust 
\cite{le96}. Such a  heating could be generated for instance by a crustquake.
In this respect, one must mention that the differential rotation between 
the superfluid and the crust generates a centrifugal buoyancy force which 
might increase the crust stresses and maybe lead to a glitch \cite{cls00}.

\section{Relativistic description}
Until very recently, 
only Newtonian theory was used in the  studies of  superfluidity
in  neutron stars, whereas, in parallel, general relativistic studies 
of neutron stars, even numerical, were based on perfect fluid matter.
The purpose of the work summarized here is to establish a bridge between 
these two approaches, taking into account both the highly relativistic 
nature of neutron stars and their  superfluid interior. We shall present  
a macroscopic formalism \cite{cl3} allowing for the average effect of vorticity 
quantisation in a rotating superfluid, which can then be extended \cite{cl98}
to describe as well superconducting fluids, such as protons in the core 
of neutron stars. The Newtonian version of  this formalism can be found in 
\cite{ml91}.
In  the next section, a simplified relativistic two-component model of neutron star, which allows for differential rotation of the superfluid
component, will be used to study oscillations of neutron stars.

\subsection{Perfect fluid in general relativity}
Let us start by recalling  the simplest case, that of the perfect fluid. 
In general relativity, a perfect fluid is characterized by 
\begin{itemize}
\item
a four-velocity 
vector $u^\mu$ (where $\mu=0,1,2,3$ is a spacetime index 
denoting  the time coordinate and the three space coordinates), which 
satisfies  the normalization $g_{\rho\sigma}u^\rho u^\sigma=-1$ where 
$g_{\rho\sigma}$ is the metric tensor describing the geometry of spacetime
(and which is used to raise or lower the indices, e.g. $u_\rho\equiv 
g_{\rho\sigma}u^\sigma$).

\item 
a particle number density scalar field $n(x^\mu)$ giving in each spacetime 
point the density of particles {\it as seen by an observer comoving with the 
fluid}.

\item 
an equation of state of the form $\rho=\rho(n)$ giving the energy density 
$\rho$ as a function of the particle number density $n$.

\end{itemize}
Note that we have chosen here the simpler case of a barotropic perfect fluid 
where the equation of state depends on only one parameter. Note also that 
one could characterize the fluid by its energy density instead of its particle 
number density, and then derive the number density by inverting the equation 
of state.

The equations of motion for the relativistic fluid will be generalizations 
of the well-known Newtonian equations of fluid mechanics, namely, 
\begin{itemize}
\item the matter conservation
equation, which can be written in the present context in the very simple 
form, 
\beq
\nabla_\mu n^\mu=0, \label{conservation}
\eeq 
where $n^\mu\equiv n u^\mu$ is the {\it particle current} and $\nabla_\mu$ 
stands for the {\it covariant derivative}.

\item the relativistic Euler equation, which can be written in the 
very condensed form 
\beq
n^\rho\nabla_{[\rho}\mu_{\sigma]}=0,  \label{euler}
\eeq
where the brackets represent antisymmetrization on the indices and where 
one has conveniently introduced the momentum covector 
$\mu_\rho\equiv \mu u_\rho$,  $\mu$ being  the relativistic {\it chemical 
potential} defined by 
\beq
 \mu={d\rho\over dn}.
\eeq
\end{itemize}
It is possible to derive directly the above equations of motion from 
a variational principle, the Lagrangian density being simply the 
energy density considered as a function of the particle number current 
$n^\mu$. Any variation of $n^\mu$ is not allowed, but only the {\it convective}
variations, which correspond to variations of the particle flow lines 
(see \cite{carter} for more details).

\subsection{Relativistic superfluid}
What will characterize a ``pure superfluid'' (i.e. a superfluid at zero 
temperature, without gas of excitations), with respect to the more general 
class of perfect fluids, is the fact that {\it its motion is locally 
irrotational}.
In the standard language, this is a consequence of the fact that the 
`superfluid velocity' can be expressed as the gradient of a quantum scalar 
phase. Strictly speaking, the standard `superfluid velocity' is in fact a 
{\it momentum} divided by a somewhat arbitrary mass. In the relativistic 
context, this will be generalized by the expression
\beq
\mu_\sigma=\hbar\nabla_\sigma\varphi, \label{superfluid}
\eeq
so that the property of irrotational flow can be written as the vanishing 
of the vorticity tensor
\beq
w_{\rho\sigma}\equiv 2\nabla_{[\rho}\mu_{\sigma]}.
\eeq

The previous condition is valid locally. However, as explained above, 
a superfluid can exhibit a non-irrotational flow at the price of being 
threaded by quantized vortices. This means that macroscopically, on distances 
bigger than the typical intervortex separation, one would like to describe 
the superfluid as well as the vortices in an average way. This can be 
done by considering the vorticity tensor $w_{\rho\sigma}$, now non-zero 
since there are vorticies, as a fundamental quantity. In fact, this tensor 
contains information about the density of vortices as well as about the 
local average direction of the vortex array. 

The idea then is to construct a generalized Lagrangian density that depends 
not only on the particle current $n^\sigma$ but also on the vorticity 
tensor $w_{\rho\sigma}$. The variations with respect to these fundamental 
variables will define the canonical momenta,
\beq
\delta\Lambda=\mu_\sigma\delta n^\sigma -{1\over 2}\lambda^{\rho\sigma} \delta
w_{\rho\sigma}.
\eeq
This can be seen as a generalization of the perfect fluid where $\Lambda=-\rho$ and only the first term is present on the right hand side.

The equations of motion   are the matter conservation equation 
(\ref{conservation}), as in the 
perfect fluid case, and a generalized Euler-type equation of the form,
\beq
w_{\mu\nu}(n^\nu-\nabla_\rho\lambda^{\rho\nu})=0.
\eeq
Of course, the expression for $\lambda^{\rho\sigma}$, which essentially 
represents  the energy density and momentum of an individual vortex, must be 
specified by resorting to a `microphysical' model of the vortices (see 
\cite{cl2} and \cite{prix} for relativistic descriptions of vortices).

\subsection{Superfluid-Superconducting mixtures}
\font\srm=cmr9
\def\eX{e^{_{\rm X}}}
\def\piX{\pi^{_{\rm X}}}
\def\wX{w^{_{\rm X}}}
\def\nX{n_{_{\rm X}}}
\def\muX{\mu^{_{\rm X}}}

It is useful to extend the previous formalism to the case where the 
superfluid particles (or Cooper pairs)  are electrically charged, like 
electrons in laboratory superconductors, in order to be able to describe 
the protons in the cores of neutron stars. 
Let us consider several species labelled by the index {\srm X}, with 
respective particle currents $\nX^{\,\rho}$ and respective 
electric charge per particle $\eX$. The total electric current is then 
given by 
 \beq
j^\rho=\sum_{\rm X}\eX \nX^{\,\rho} \, .
\eeq

Proceeding as before, In order to obtain the equations of motion for the 
global system of fluids, 
one  uses a variational principle based on a Lagrangian density
${\cal L}$ that will be the sum of three contributions:
\begin{itemize}
\item a ``matter''
contribution $\Lambda_{\rm M}$, which depends only on the `hydrodynamical' 
part of the system, and which is a function only of the particle currents 
$\nX^{\,\rho}$, or more exactly of all scalar combinations obtained by their 
mutual contractions,

\item 

an electromagnetic interaction term depending on the electromagnetic gauge
form  $A_\rho$, 
\beq  
\Lambda_{\rm I}= j^\rho A_\rho \, ,
\eeq

\item 
an electromagnetic field
contribution 
 \beq
\Lambda_{\rm F}={1\over 16\pi} F_{\rho\sigma}F^{\sigma\rho}\, .
\eeq
where the electromagnetic field tensor $ F_{\rho\sigma}$ is defined in terms 
of the gauge form by the usual expression 
$F_{\rho\sigma}=2\nabla_{[\rho}A_{\sigma]}$. 

\end{itemize}

Considering the variations of ${\cal L}= \Lambda_{\rm M}+\Lambda_{\rm F}+
\Lambda_{\rm I}$ with respect to the particle currents $\nX^{\,\rho}$
 naturally suggests to define the  canonical momentum covectors
\beq
\piX_{\ \rho}=\muX_{\ \rho}+\eX A_\rho\, 
\eeq
which are the sum of a pure `hydrodynamical' (or `chemical') part and of 
the electromagnetic gauge form weighted by the electric charge of the species.
From this, it is convenient to define generalized vorticity tensors defined 
by 
 \beq
\wX_{\,\rho\sigma}=2\nabla_{[\rho}\piX{_{\!\sigma]}}
=2\nabla_{[\rho}\muX{_{\!\sigma]}}+\eX F_{\rho\sigma}\, .
 \eeq

The equations of motion for the system consist of the separate matter 
conservation equation for each species (one can generalize to allow for 
chemical reactions between various species \cite{lsc98}),
\beq
\nabla_\rho \,\nX^{\,\rho}=0 \, ,
\eeq
and of Euler-type equations, which can be written in the very compact form
\beq
\nX^{\,\sigma}\wX_{\,\sigma\rho}=0 \, .
\eeq
Note that in the case of a charged component, this equation contains the 
electromagnetic field tensor and thus automatically includes for instance 
the Lorentz force exerted on the fluid. 

It is also useful to write for the system under investigation 
the associated total stress-energy-momentum tensor, which appears  on 
the right hand side of the Einstein's equations,
\beq
R_{\mu\nu}-{1\over 2}R g_{\mu\nu} = 8\pi G T_{\mu\nu}, \label{einstein}
\eeq
if one needs to determine the spacetime metric. 
Once again, the variational principle is very useful because variation 
of the Lagrangian density with respect to the metric directly yields 
\beq
T^{\rho\sigma}=T_{\rm M}^{\ \rho\sigma}+T_{\rm F}^{\ \rho\sigma}\, ,
\eeq
where the part derived  from the material Lagrangian density 
contribution $\Lambda_{\rm M}$ is given by
\beq
T_{{\rm M}\ \, \sigma}^{\ \rho}=\sum_{\rm X}\nX^{\,\rho}\muX_{\,\sigma}+
s\Theta u^\rho u_\sigma +\Psi_{\rm M} g^\rho_{\,\sigma}\, ,\hskip 1 cm
\Psi_{\rm M}=\Lambda_{\rm M}-\sum_{\rm X}\nX^{\,\sigma}\muX_{\,\sigma}+s\Theta\, ,
\eeq
while the electromagnetic contribution has the usual Maxwellian form 
 \beq
T_{{\rm F}\ \, \sigma}^{\ \rho}={1\over 4\pi}\Big( F^{\nu\rho} F_{\nu\sigma}
-{1\over 4}F^{\mu\nu}F_{\mu\nu} g^\rho_{\,\sigma}\Big)\, .
 \eeq

The above relations are valid for ordinary perfectly conducting fluids. 
The condition characterizing superconductors, generalizing the superfluid 
condition (\ref{superfluid}), is that the {\it generalized momentum} should 
be the gradient of a quantum phase, 
\beq
\piX_{\ \rho}
\equiv  \muX_{\ \rho}+\eX_\rho=\hbar \nabla_\rho\,\varphi^{_{\rm X}}\, ,
\eeq
so that superconductors (including the case of superfluids) are characterized 
{\it locally} by a vanishing generalized vorticity tensor. 
Of course, it is then possible, in analogy with the treatment of the previous 
subsection,  to extend the formalism in order to 
describe on average a superfluid-superconducting mixture threaded by 
superfluid vortices as well as magnetic flux tubes. 

\section{Relativistic  neutron stars}
Let us now see how one can construct a concrete model of neutron star in 
a general relativistic framework. To simplify, we will assume that the neutron 
star is made of simply two components, each being treated as a  perfect fluid.
First, we will consider the case of a static, i.e. nonrotating, neutron star.
Then we will consider small, i.e. linear, oscillations of the two-component 
neutron star.

Let us call, loosely, the component corresponding to the free neutron 
superfluid the ``neutron'' component and the component corresponding to the 
normal fluid the ``proton'' component. Their respective particle currents
will be denoted $n^\rho=nu^\rho$ and $p^\rho=pv^\rho$, where $u^\rho$ and 
$v^\rho$ are unit four-velocities, which are not aligned if the two components
are not comoving. Their equations of motion consist of two separate 
conservation equations of the type (\ref{conservation}) and of two Euler 
equations of the type (\ref{euler}). Here, the two components are treated 
as ordinary perfect fluids and one ignores the refinements due to the presence 
of vortices in the superfluid component. 

\subsection{Static neutron star}
We first consider the configuration corresponding to a non-rotating 
neutron star, which is thus {\it static} and {\it spherically symmetric}
and for which the  metric can  be written in the form 
\beq
ds^2=g_{\mu\nu}dx^\mu dx^\nu=-e^{\nu(r)}dt^2+e^{\lambda(r)}dr^2
+r^2\left(d\theta^2+\sin^2\theta d\phi^2\right).
\eeq
The particle currents, since the two fluids are motionless, are necessarily 
of the form 
\beq
n^\mu=n\bar u^\mu, \quad 
p^\mu=p\bar u^\mu, \quad
\bar u^\mu=\{u^0,0,0,0\}.
\eeq
Writing Einstein's equations with the above ans\"atze, one obtains the 
well-known TOV (Tolman-Oppenheimer-Volkov) equations:
\beq
{dP\over dr}=- G{(\rho(r)+P(r))(m(r)+4\pi r^3 P(r))\over r^2(1-2 Gm(r)/r)}=
-{1\over 2}(\rho(r)+P(r)){d\nu\over dr},
\eeq
and
\beq
e^{\lambda(r)}=\left(1-2{m(r)\over r}\right)^{-1},
\eeq
with $m(r)\equiv 4\pi G\int_0^r dr' \rho(r'){r'}^2$.
The mass of the star is given by $M=m(r)$, where 
$R$ is the star radius defined by the condition that the pressure vanishes.

\subsection{Oscillations of superfluid neutron stars}

Oscillations of relativistic stars have been studied by many authors in the 
case of a {\it single} perfect fluid. Here, we summarize the first 
investigation for a two-component relativistic star \cite{cll}, thus 
generalizing two previous studies of two-component Newtonian stars 
\cite{newton}. 

In a relativistic approach, one must consider not only the perturbations 
of the quantities describing the matter, for example the particle density and 
the velocity, but also the perturbations of the spacetime metric $g_{\mu\nu}$.
After decomposition of the perturbations into spherical harmonics,
labelled by $l$ and $m$,  it is 
convenient to distinguish even-parity (or polar) perturbations and 
odd-parity (or axial) modes depending on their transformation under parity. 
We will restrict ourselves here to the case of  even-parity perturbations,
which is the most interesting since scalar quantities, such as the matter 
densities, give only this  type of  perturbations. Also, one needs to 
study only $m=0$ perturbations, since there is a degeneracy in $m$ for each 
given $l$ because of the spherical symmetry of the background.

Choosing a specific gauge, one will consider metric perturbations of the 
form 
\begin{eqnarray}
&\delta g_{00}=-e^\nu r^l H_0 e^{i\omega t}P_l(\theta), 
&\delta g_{0r}=\delta g_{r0}= -i\omega r^{l+1} H_1 e^{i\omega t}P_l(\theta),\cr
&\delta g_{rr}= -e^\lambda r^l H_2 e^{i\omega t}P_l(\theta), 
&\delta g_{\theta \theta}=\delta g_{\phi\phi}/\sin^2\theta=
- r^{l+2} K e^{i\omega t}P_l(\theta),
\end{eqnarray}
where $P_l$ stands for the Legendre polynomial of order $l$.
The matter perturbations will be described, in a Lagrangian way, by the
matter displacements, 
\begin{eqnarray}
\xi_n^r=&r^{l-1}e^{-\lambda/2}W_n e^{i\omega t}P_l(\theta), \quad
\xi_n^\theta=&- r^{l-2}V_n e^{i\omega t}\partial_\theta P_l(\theta),\cr
 \xi_p^r=&r^{l-1}e^{-\lambda/2}W_p e^{i\omega t}P_l(\theta), \quad
\xi_p^\theta=&- r^{l-2}V_p e^{i\omega t}\partial_\theta P_l(\theta),
\end{eqnarray}
from which one can compute the perturbed velocities, by taking  the time 
derivative or the variations $\delta n$ and $\delta p$ by using the 
perturbed conservation equations. 
Inserting the above expressions for the perturbations into the perturbed 
Einstein's equations as well as the perturbed Euler equations, one ends 
up with a system of linear equations, consisting of two contraints (one 
is $H_2=H_0$, the other  expresses  $H_0$ in terms of the other 
perturbations) and of first order differential equations of the form 
\beq
{dY\over dr}= Q_{l,\omega}Y, 
\eeq
where $Y$ is 6-dimensional column vector containing $H_1$, $K$, $W_n$, $V_n$, 
$W_p$ and $V_p$, and $Q_{l,\omega}$ is a $6\times 6$ matrix with
$r$-dependent  coefficients
that depend only on the background configuration, as well as on $l$ and 
$\omega$.
Considering the boundary conditions at the center and at the surface of the 
star, this system can be solved, up to a global amplitude, for {\it any value} 
of $\omega$. The physically relevant modes, however, also called quasi-normal 
modes, correspond  the specific values 
of $\omega$ for which the metric outside 
the star represents {\it only outgoing gravitational waves}.

A numerical investigation for a very crude, and unrealistic, model of two 
independent polytropes (with different adiabatic indices) has shown that 
the two-component star will exhibit new modes, {\it superfluid} modes, 
which are specific of the existence of two components since the two fluids 
are counter-moving for these modes.

\end{document}